\documentclass[preprint,11pt]{aastex}

\slugcomment{ApJ, in press} 

\begin{document}

\title{Light Curves of Rapidly Rotating Neutron Stars} 

\author{Timothy M. Braje and Roger W. Romani} \affil{Dept. of Physics,
Stanford University, Stanford, CA 94305-4060}
\email{timb@astro.stanford.edu; rwr@astro.stanford.edu} 

\and 

\author{Kevin P.  Rauch} \affil{Dept. of Astronomy, University of
Maryland, College Park, MD 20742-2421} \email{rauch@astro.umd.edu} 

\begin{abstract}
        We consider the effect of rapid rotation on the light curves
of neutron stars with hot polar caps.  For $P \approx 3$ms spin
periods, the pulse fractions can be as much as an order of magnitude
larger than with simple slowly-rotating (Schwarzschild)
estimates. Doppler boosting, in particular, leads to characteristic
distortion and ``soft lags'' in the pulse profiles, which are easily
measurable in light curves with moderate energy resolution. With 
$\sim 10^5$ photons it should also be possible to isolate the more subtle
distortions of light travel time variations and frame dragging.
Detailed analysis of high quality millisecond pulsar data from
upcoming X-ray missions must include these effects. 

\end{abstract}


\section{Introduction}

	Many neutron star models have thermal surface emission
concentrated at magnetic polar caps, which are heated by accretion
columns, by precipitating magnetospheric particles, or by anisotropy
in the thermal conductivity and heat flow from the interior.  A number
of authors have computed light curves from such heated polar caps by
assuming a stationary dipolar magnetic field and noting that
gravitational focusing of the photons can have a substantial effect on
the pulse shape (see, e.g. Riffert \& M\'{e}sz\'{a}ros~1988; Zavlin,
Shibanov, \& Pavlov~1995).  It has also been suggested that resonant
scattering on magnetospheric $e^\pm$ may modify the thermal light
curves \citep{raj97,zhu97}.  Recently, pulsed X-ray emission,
apparently from the stellar surface, has been detected from a number
of very short period neutron stars, including the 5.75ms radio pulsar
J0437-4715 \citep{zav98} and the 2.49ms LMXB binary J1808.4-3658
\citep{cui98,wij98}.  For such objects, existing static field and
metric approximations may be inadequate. 

	In this paper we model the effects of rapid rotation on the
the surface emission from polar caps.  The most prominent effects are
associated with simple Doppler boosting at the surface
\citep{for99,str98}. We also model the perturbations associated with
the varying time delays of the curved light paths and the pulse
distortions resulting from frame-dragging (labeled here as ``Kerr'')
effects on the space-time metric. 
Previously, \cite{che89} noticed the above effects.  They modeled the
neutron star as a Kerr black hole, and computed values for the star's
angular momentum while assuming that the neutron star's surface is
stationary in the LNRF (see below).  Because of this, they arrive at
unphysical values for the angular momentum and only include Doppler
shifts due to frame-dragging.  We use more realistic values of the
frame-dragging angular velocity and include the dominant Doppler
boosts caused by the motion of the star's surface in the LNRF.
These effects are subtle, but with the increased X-ray sensitivity of
{\it AXAF}, {\it XMM}, and possibly {\it USA} careful observations may
uncover signatures of rotational perturbations in high statistics
energy-resolved light curves of the shortest-period pulsars. The
effects will be even more prominent when emission arises from higher
altitudes, as is likely for the non-thermal radio and high energy
pulsations found in a number of millisecond pulsars (e.g.  PSR
B1821-24, PSR J0218+4232).  Resonant scattering perturbations can also
show much larger Doppler and Kerr effects since the flux is
re-directed at several $R_{*}$ and sent back towards the star.
Modeling these sources additionally requires a detailed treatment of
the retarded (swept-back) $B$ field. 

\section{The Model} \label{model}

	We start by emitting photons from heated polar caps, whose
boundary is defined by the last ``open'' field lines. For this
paper we adopt the traditional static field, aligned-rotator estimate 
of the cap size and use units where $G=c=1$ 
\begin{equation} \label{cap}
	\sin\theta_{\rm{cap}} = \sqrt{ R_{*}\Omega_{*} }, 
\end{equation}
where $ \theta_{\rm{cap}}$ is the half-angular size of the polar cap, 
$ R_{*} $ is the radius of the star, $ \Omega_{*} $ is the angular
velocity of the star, Our fiducial model has both caps heated and, for
the old, cold neutron stars which have been ``recycled'', or spun up
to short periods, we neglect emission from the rest of the star
surface.  For a rough approximation, we emit a blackbody spectrum,
which will suffer simple interstellar extinction before reaching the
observer.  The surface emission is assumed to be moderately beamed 
$I_\nu(\theta_{\rm{n}} ) \propto {\cos}(\theta_{\rm{n}} )$ 
to mimic the effect of limb darkening. We have also considered more
complicated surface temperature distributions.  Fits to real data,
seeking to find rapid rotation effects in pulsar light curves, will
need to propagate emission from a more realistic thermal spectrum
including composition effects, accurate limb darkening, and surface
$T$ variation. The present sums, however, suffice to illustrate the
size and characteristic shape of rotation-induced light curve
perturbations and will be useful as a guide to when such effects need
to be considered. 

	Formally, our use of Eq.~\ref{cap} is inconsistent.  We have
computed the cap distortions induced by the rotational sweep-back of
the field lines for a non-aligned dipole (e.g. Deutsch~1955; Romani \&
Yadigaroglu~1995) including near-surface modifications due to
space-time curvature in a Schwarzschild metric
\citep{mus97,pra97,lin75}.  By generalizing the computation of the
field structure of a point dipole in the Schwarzschild metric, we have
obtained an expression for a point, rotating dipole in a Schwarzschild
space-time that goes to the static Schwarzschild case when $\Omega_{*}
= 0$, goes to the rotating point dipole solution when $r$ gets large,
and satisfies Maxwell's equations in the limits $M \rightarrow 0$ and
$\Omega_{*} \rightarrow 0$, where $M$ is the mass of the pulsar.   We
can then use this expression to solve for the field lines that are
tangent to the light cylinder and define the boundary of the open
zone.  These lines, when traced back to the surface, result in an
irregular shape for the field line foot points and cap boundary.  We
find that the characteristic cap size is quite close to the simple
static estimate for $P \ge 1$ms.  After the inclusion of gravitational
focusing, the changed cap shape has negligible effect on the light
curve; however, when higher altitude emission or scattering are
considered, the structure of the relativistic swept-back field is
quite important.  We defer discussion of these more complex light
curves (T.M. Braje \& R.W. Romani, in preparation), focusing here on direct
surface emission. 

	Our computation is a Monte Carlo simulation, using the method
developed to model resonant scattering perturbations by
\citet{raj97}. We have extended the photon propagation code to include
frame-dragging effects and have modified the transport to follow
multiple energy slices of the emitted spectrum, tracing spectral
variation through the pulse.  We assume a $1.4 M_{\sun}$, $10$km
($4.84\rm{GM}/\rm{c}^2$) radius star, and a fiducial spin period
$P=3$ms which gives a cap size of half-angle $15\arcdeg$. 

\subsection{Boosts} \label{boosts}

	The initial photon directions are drawn in the surface frame
with angle $\theta_{\rm{n}}$ distributed relative to the local normal
with probability proportional to
$ \sin\theta_{\rm{n}}\cos\theta_{\rm{n}} $, and 
uniformly distributed
in $\phi$.  Although magnetic anisotropies in the surface opacity may
break the azimuthal symmetry for higher field pulsars \citep{pav94,rrm97}, this
is unlikely to be important for low $B$, short period neutron
stars. These angles give an initial four-momentum of the photon, 
\begin{equation} \label{initmomentum}
	p^{\nu}_{\rm{init}} = E_{\rm{surf}}
			(1,\sin \theta_{\rm{n}} \cos \phi, \sin
			\theta_{\rm{n}} \sin \phi, \cos \theta_{\rm{n}}). 
\end{equation}

	To start the photon trajectories, we must relate a frame
co-rotating with the surface of the star, where the photons are
emitted with some prescribed spectrum and angular distribution, to a
locally Lorentz coordinate frame (Locally Non-Rotating Reference
Frame, hereafter the LNRF; e.g., Bardeen, Press, \& Teukolsky~1972).
The surface velocity of the pulsar, as measured in the LNRF, is 
\begin{equation} \label{boostvel}
	v_{\hat{\phi}} = ( \Omega_{*}-\Omega ) \left [ \frac{-g_{\phi
		\phi}}{g_{tt}+\Omega g_{t\phi}} \right ]^{\frac{1}{2}} 
\end{equation}
where the stellar angular frequency is $\Omega_{*}$ and the reference
frame rotates at $ \Omega = -g_{t \phi}/g_{\phi \phi} $. 
We make a local Lorentz boost to the LNRF 
\begin{equation}
	p^{\mu}_{\rm{final}} =
	\Lambda^{\mu}_{\nu}(v_{\hat{\phi}})p^{\nu}_{\rm{init}}, 
\end{equation}
choosing the coordinate system so that the y-axis points along the
direction of motion, giving 
\begin{equation}
	\Lambda^{\mu}_{\nu}(v_{\hat{\phi}}) = \left(
		\begin{array}{cccc} \gamma & 0 & -\gamma
		v_{\hat{\phi}} & 0 \\ 0 & 1 & 0 & 0 \\ -\gamma
		v_{\hat{\phi}} & 0 & \gamma & 0 \\ 0 & 0 & 0 & 1 \\
		\end{array} \right). 
\end{equation}

Finally, we convert the photon four-momentum to Boyer-Lindquist
coordinates, using the transformations: 
\begin{equation}
   \begin{array}{ccl} X^{\hat{t}} & = & (-g_{tt}-\Omega g_{t
	\phi})^{1/2} X^{t} \\ X^{\hat{r}} & = & (g_{rr})^{1/2}X^{r} \\
	X^{\hat{\theta}} & = & (g_{\theta \theta})^{1/2} X^{\theta} \\
	X^{\hat{\phi}} & = & (g_{\phi \phi})^{1/2} [X^{\phi}-\Omega
	X^{t}] \\ \end{array} 
\end{equation}
where $X^{\mu}$ represents any four-vector, the left-hand quantities
are in the orthonormal LNRF and the right-hand quantities are measured
in the Boyer-Lindquist frame. 

These initial directions are propagated to large radius 
(8$R_{*}$).  The energy boost factor is used to re-weight the surface
spectral distribution in the final photon spectrum. 
We choose a temperature as measured at the surface of the star of
$1.3 \times 10^6$, so that after the Schwarzschild gravitational
redshift, the temperature at infinity is $10^6$K.

\subsection{Time Delay} \label{time delay}

	A proper apportionment of photons among final rotational phase
(time) bins must include time delay effects arising from both the
variation in the curved path length traveled to ``infinity'' and from
the variation in the gravitational time delays experienced by the null
geodesics near the star. In practice, we integrate to a fiducial large
radius 8$R_*$ beyond which flat space propagation, and a constant
incremental light travel time, takes us to the observer.  The net
effect is that the phase coordinate on the two-dimensional sky-map of
a photon is corrected by 
\begin{equation}
	\phi_{\gamma} \rightarrow \phi_{\gamma} - \Omega_{*}t 
\end{equation}
where $t$ is the (coordinate) time 
integrated from the surface to the fiducial radius. 

This delay will have its strongest effect at the tails of the light
curves, where the photons have both traveled the farthest to reach
their final destination and spent the most time near the star,
suffering bending and time delay in the gravitational field. 

\subsection{Kerr Propagation} \label{kerr}

	Frame dragging effects around a rotating neutron star will
depend on the detailed mass distribution, and hence on the equation of
state (EOS) and the spin period.  Since neutron stars are moderately
centrally condensed and the surface emission starts propagation at 
$ r \geq 3M $, we can adopt a ``Roche approximation'', assuming all mass
is 
near the origin.
This allows us to obtain an approximate set of trajectories by using
null geodesics of the Kerr metric. For our chosen pulsar period, mass
and radius, we estimate appropriate values for $a ( =J/M )$ using the
numerical models calculated by \citet{coo94}.  We adopt $a =
0.47/P({\rm ms})$ measured in units of M as appropriate to their
moderately soft ``FPS'' model at $M=1.4 M_{\sun}$ and low
$R_*$.  We retain, however, the fiducial $10$km radius estimates; note
that this is conservative in the sense that Doppler effects at the
true $\sim 12$km radius can be even larger.  Note also that $a \propto
\Omega_{*}$ breaks down at $P<1$ms as the star is rotationally
flattened, giving even larger $R_*$ \citep{coo94}.

Because of certain numerical integration advantages, we use the Kerr
metric in a slightly modified form \citep{rau94}: 
\begin{equation}
   \begin{array}{ccl} ds^{2} & = & - dt^{2}\left( 1 -
	\frac{2Mu}{\tilde{\rho}^{2} } \right) + du^{2} \frac{
	\tilde{\rho}^{2} }{u^{4} (1-2Mu+a^{2}u^{2}) } + d \mu^{2}
	\frac{ \tilde{\rho}^{2} }{u^{2}(1- \mu^{2})} \\ \cr & & + d
	\phi^{2} \frac{ (1- \mu^{2})(u^{-2}+ a^{2} [ \tilde{\rho}^{2}
	+ \mu^{2} + 2Mu(1- \mu^{2})])} {\tilde{\rho}^{2}} - 2d \phi dt
	\frac{2aMu(1- \mu^{2})}{\tilde{\rho}^{2}} \end{array} 
\end{equation}
where \( \mu=\cos\theta \), \(u = r^{-1} \), and $ \tilde{\rho}^{2} =
1+a^{2} \mu^{2} u^{2} $.  The equations of motion are 
\begin{equation}
  \begin{array}{ccccl} p^{t} & = & \frac{dt}{Ed\lambda} & = &
	\rho^{-2} \left[
	-a(a(1-\mu^{2})-l)+\frac{u^{-2}+a^{2}}{\Delta}
	(u^{-2}+a^{2}-al) \right] \\ \cr p^{u} & = &
	\frac{du}{Ed\lambda} & = &
	u_{\rm{sgn}}\rho^{-2}\sqrt{1+(a^{2}-q^{2}-l^{2})u^{2} +2(
	(a-l)^{2}+q^{2})u^{3}-a^{2}q^{2}u^{4}} \\ \cr p^{\mu} & = &
	\frac{d\mu}{Ed\lambda} & = &
	\mu_{\rm{sgn}}\rho^{-2}\sqrt{q^{2}+(a^{2}-q^{2}-l^{2})\mu^{2}
	-a^{2}\mu^{4}} \\ \cr p^{\phi} & = & \frac{d\phi}{Ed\lambda} &
	= & \rho^{-2} \left[ -a+\frac{l}{1-\mu^{2}} +
	\frac{a}{\Delta}(u^{-2}+a^{2}-al) \right] \\ \end{array} 
\end{equation}
where $ \rho^{2}=u^{-2}+a^2\mu^{2} $; $ \Delta = u^{-2}-2Mu^{-1}+a^{2}
$; and $u_{\rm{sgn}}$ and $\mu_{\rm{sgn}}$ are signs that are dependent on the
initial direction of the photon. For the initial four-momentum
$p^{\nu}$ of our photon in the Boyer-Lindquist frame
(Eq.~\ref{initmomentum}), we can compute the constants of motion ($l$
is the z component of angular momentum and $q^2$ is Carter's constant,
both normalized in units of the energy at infinity, $E=-p_{t}$) as 
\begin{eqnarray}
	l = L/E & = & -p_{\phi}/p_{t} \\ q^{2} = {\mathcal{Q}}/E^2 & =
	& ( p_\theta/p_t )^2 + \mu^2 [ -a^2+l^2/(1-\mu^2) ] 
\end{eqnarray}
where, as usual, $ p_{\nu} = g_{\nu \sigma}p^{\sigma}$.  In order to
speed up the computations, we use an elliptic integral reduction of
the geodesics presented in \citet{rau94}. 

\section{Numerical Results}

	For our fiducial $\alpha=60\arcdeg$, $P=3$ms case, we have
produced high-precision light curves by drawing $\sim10^8$ photons and
mapping the phase distribution in each of 10 energy bands.  The
computations use simple Schwarzschild propagations (to match the
efforts of earlier authors), as well as progressively adding the
Doppler boosts, time delays and frame-dragging propagation effects
described in Sec.~\ref{model}. 

\begin{figure}[htb]
\epsscale{.8}
\plotone{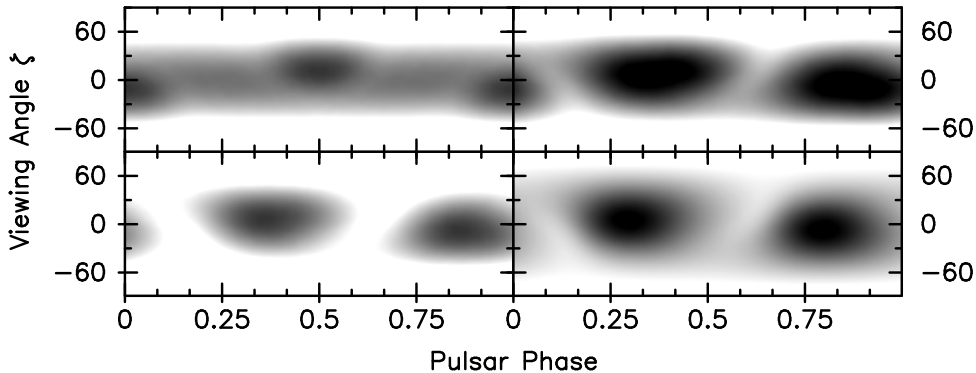}
\caption{All-sky maps of polar cap emission.  Parameters:
$\alpha=60\arcdeg$, $P=3$ms, two caps of $T_{\rm{surf}}=1.3 \times 10^6$~K
after
interstellar absorption.  \emph{Upper Left:} $P \approx 0$
(Schwarzschild), all energies.  \emph{Lower Left:} All rotational
effects, all energies, shown at the same intensity level as Schwarzschild. 
\emph{Upper Right:} All effects, low energy
band ($0.07-0.10$~keV).  \emph{Lower Right:} All effects, high energy
band ($0.8-1.1$~keV).  Note the significant distortions and phase
shifts.\label{skymap}} 
\end{figure}

	The results are displayed as sky-maps (Fig.~\ref{skymap}) with the
intensity of radiation plotted as a function of viewing angle $\zeta$
and pulsar phase $\phi$.  We produce light curves from such images by
taking a slice along a particular $\zeta$. 

In Fig.~\ref{skymap}, the left panels compare the maps produced by
Schwarzschild alone (upper) and the combined rotational effects
(lower). Several important differences emerge.  In the Schwarzschild
map, slightly lighter (fainter) bands border each cap; these are the
``shadows'' of the opposite pole cast by the star and only partially
filled by the facing polar cap.  With boosting, time delay and frame
dragging, the beam of the cap emission is shifted forward and the
simple symmetry around the magnetic axis is broken, so these circular
shadows do not appear.  The right panels show the variation in the
sky-map expected with photon energy. For a $T_{\rm{surf}} = 1.3 \times 10^6$K,
these maps represent the energy bands (at infinity)
$0.07-0.10$~keV (upper) and $0.8-1.1$~keV
(lower), respectively. The dramatic enhancement of the hard emission
at the leading edge of the pulse is clear, showing the Doppler boost's
dominant contribution to the pulse shift. 

\begin{figure}[htb]
\epsscale{.8}
\plotone{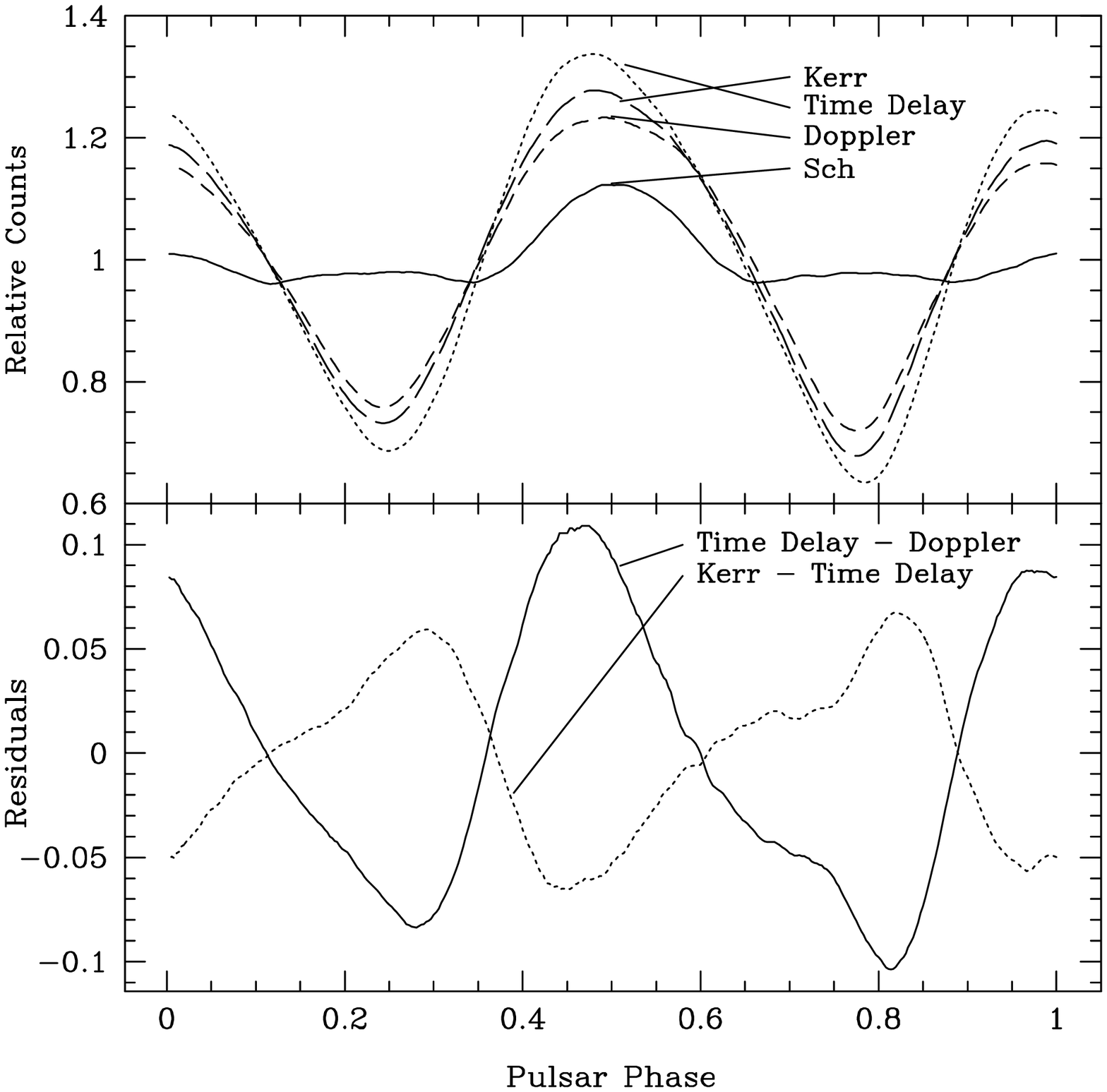}
\caption{Rotational effects on polar light curves
($P=3$ms, $\alpha=60\arcdeg$, $\zeta=80\arcdeg$, cap radius
$15\arcdeg$).  Curves have been shifted to common phase of a best-fit
$P/2$ sinusoid and normalized to a mean count~rate per bin of unity.
\emph{Above:} Effects of including rotational perturbations.
\emph{Below:} Residuals from neglecting time delay or
Kerr. \label{all}} 
\end{figure}

	Simple light curves show more clearly the relative
contributions to the pulse shape distortions. Since an observer will
only obtain one light curve ($\alpha=60\arcdeg$, $\zeta=80\arcdeg$ in
this case), the effects of the different rotational perturbations can
only be isolated by looking at shape distortions, not phase shifts.
Accordingly, we shift the light curves by the phase of the
best-fitting, two-cycle sinusoid (at $P/2$) to allow direct
comparisons (Fig.~\ref{all}). 

	As noted by earlier authors, gravitational bending in the
Schwarzschild metric greatly decreases the pulse fraction from hot
polar caps, by spreading the beam. Also, for stars with two hot caps
and masses and radii similar to the fiducial values chosen here,
gravitational focusing causes the pulse from one pole to fall in the
shadow of the opposite pole, further decreasing the pulse amplitude.
Rotational effects break this degeneracy and increase the pulse
fraction even for a symmetric two-pole model by a factor of $\sim 7$
at $P=3$ms. 

The lower panel of Fig.~\ref{all} indicates the perturbations in the
model light curves (after shifting to the best match sinusoid) which
result when the propagation time delay effects are ignored.  For
$P=3$ms these produce 8\% distortions in the light curve. If ``Kerr''
propagation effects are ignored the errors are at the 1\% level.

The pulse fraction actually decreases in going from the ``time delay''
case to the ``Kerr'' case.  This effect arises because, in the
``Kerr'' case, the LNRF has some finite angular velocity with respect
to infinity.  This angular velocity effectively reduces the initial
Doppler boost of the photon from the pulsar's frame to the LNRF.  Of
course, there is a metric-induced Doppler shift due to the angular
velocity of the LNRF (cf. t component of Eq.~6), as noted by \cite{che89},
but this is a much
smaller effect than the true Doppler shift, reducing the pulse
fraction of the ``Kerr'' light curve.

\begin{figure}[!hbt]
\epsscale{.8}
\plotone{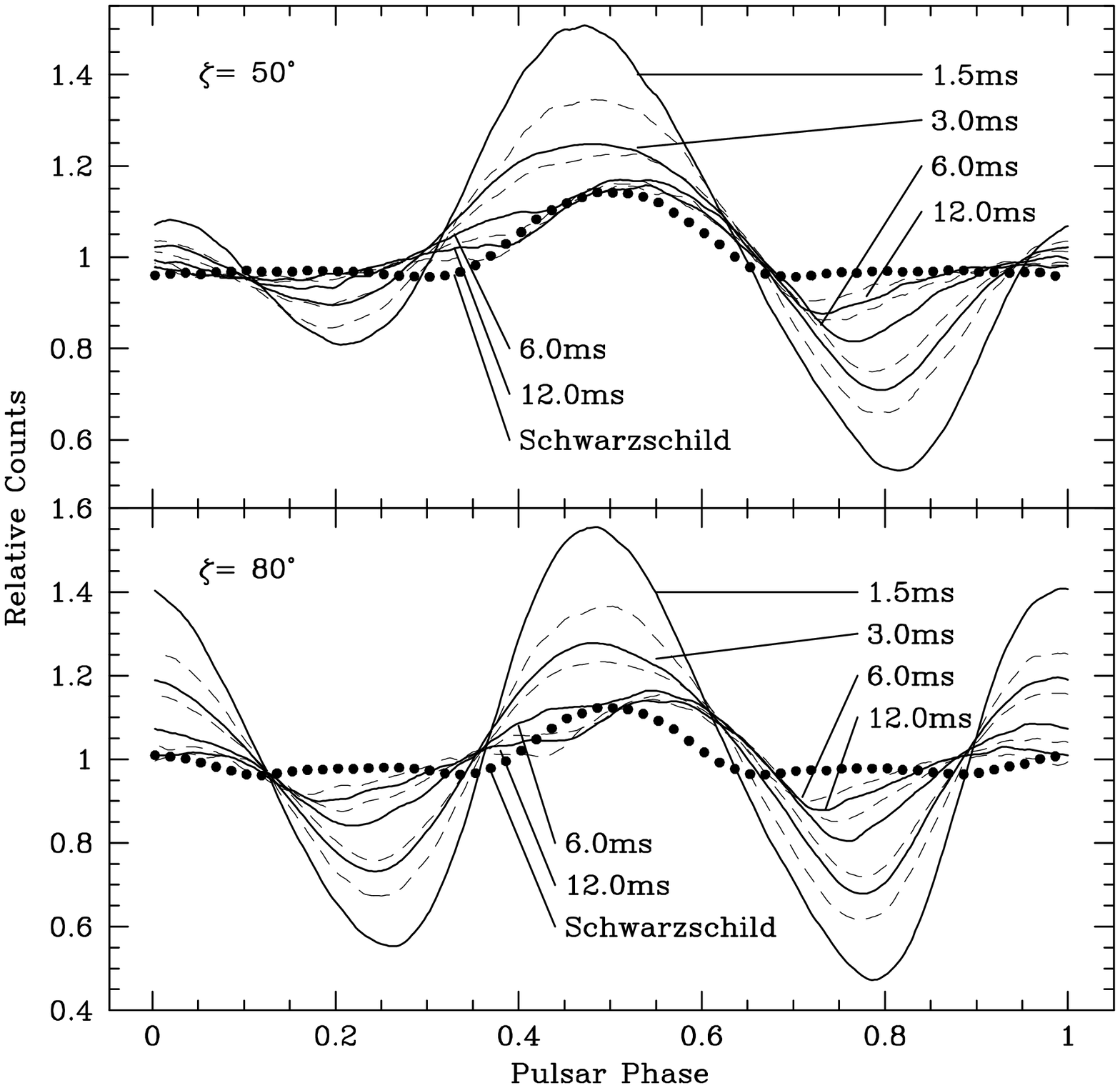}
\caption{ Period dependence of light curve distortions
($\alpha=60\arcdeg$).  Full line: All effects.  Dashed line: Doppler
boosts only.  Points: Reference $P \approx 0$ curve.  \emph{Above:}
$\zeta=50\arcdeg$.  \emph{Below:} $\zeta=80\arcdeg$. 
\label{oper}}
\end{figure}

	The period dependencies of the pulse-shape distortions are
summarized in Fig.~\ref{oper} and Table~\ref{tbl-1}.  We see that even
at modest spin frequencies the pulse fraction increases significantly
when Doppler boosts are considered.  Note that we have assumed a
relatively compact star, and surface velocities and Doppler
distortions will be even larger for a stiff EOS. 
For example, using the more realistic radius of $\sim12$km for the FPS
EOS will increase Doppler boosts by $\sim20\%$.
By $P \sim 3$ms frame dragging and time delay effects are quite
substantial.  In Table~\ref{tbl-1} we summarize the pulse fraction
(Pulsed counts/Total counts) for Doppler boosted light curves and
light curves computed with all rotational effects. These estimates
suggest that frame dragging might be measurable below $P \sim 3$ms.
In detail, the pulse fractions depend on the $\zeta$ chosen, so we
have included them for two different viewing angles. 

\begin{deluxetable}{lccccccc}
\small
\tablecaption{Sensitivity to Rapid Rotation Effects. \label{tbl-1}} 
\tablewidth{0pt} 
\tablehead{ \colhead{Statistic} & \colhead{Model} &
\colhead{$\zeta$} & \colhead{Sch} & \colhead{12ms} & \colhead{6ms} &
\colhead{3ms} & \colhead{1.5ms} } 

\startdata Pulse & Dop / Kerr & $50\arcdeg$ & 
4.3\% & 9.6\% / 12\% & 14\% / 18\% & 25\% / 30\% & 34\% / 47\% \\ 
Fraction & Dop / Kerr & $80\arcdeg$ & 
4.0\% & 10\%  / 12\% & 15\% / 20\% & 28\% / 32\% & 38\% / 53\% \\ 
\cr 
$N_{\gamma}$ & Energy shift & $50\arcdeg$ & &
$\sim1\times10^6$ & $\sim3\times10^5$ & 
$\sim6\times10^4$ & $\sim3\times10^4$ \\ 
$N_{\gamma}$ & Energy shift & $80\arcdeg$ & &
$\sim2\times10^6$ & $\sim4\times10^5$ & 
$\sim6\times10^4$ & $\sim4\times10^4$ \\ 
\cr 
$N_{\gamma}$ & Sch \emph{vs.} Dop & $50\arcdeg$ & & 
$\sim6\times10^4$ & $\sim1\times10^4$ & 
$\sim3\times10^3$ & $\sim1\times10^3$ \\ 
$N_{\gamma}$ & Sch \emph{vs.} Dop & $80\arcdeg$ & & 
$\sim7\times10^4$ & $\sim1\times10^4$ &
$\sim3\times10^3$ & $\sim1\times10^3$ \\ 
\cr
$N_{\gamma}$ & Dop \emph{vs.} Kerr & $50\arcdeg$ & & $\sim4\times10^5$ &
$\sim2\times10^5$ & $\sim7\times10^4$ & $\sim9\times10^3$ \\
$N_{\gamma}$ & Dop \emph{vs.} Kerr & $80\arcdeg$ & & $\sim4\times10^5$ &
$\sim9\times10^4$ & $\sim3\times10^4$ & $\sim5\times10^3$ \\
\enddata 
\end{deluxetable}

\begin{figure}[!ht]
\epsscale{.8}
\plotone{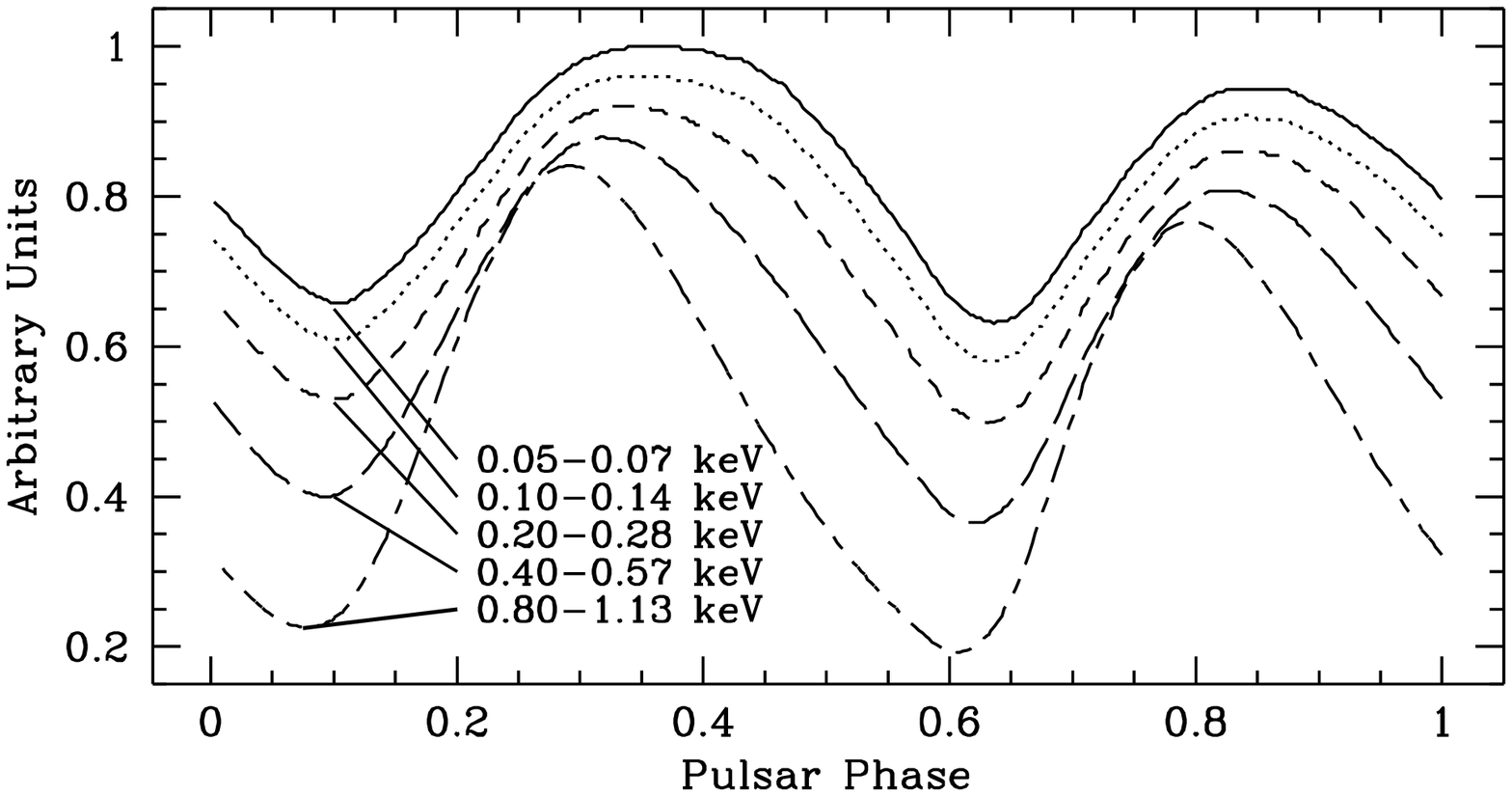}
\caption{Energy dependence on light curve shape,
showing Doppler boosting.  Arbitrary amplitude normalization. 
\label{energy}} 
\end{figure}

	Of course, by using energy cuts we can obtain different
weighting of our single line of sight across the pulsar surface.  In
Fig.~\ref{energy} we plot normalized light curves in different energy
bands for $P=3$ms.  The shifts here are inherently observable since we
have absolute phase information of each energy band with respect to
the others.  Notice that the total shift between the lowest and
highest bands is on the order of $15\arcdeg$. In principle, the energy
variations allow us to disentangle effects of the Doppler boosting
from variations in the underlying star temperature.  Sharp spectral
features, e.g.  absorption lines and edges, will make this task much
easier. 

\section{Discussion and Conclusions}

	With sufficient count statistics we might hope to isolate the
various contributions to the rotational perturbations of the pulsar
light curves.  To gauge the observational precision required, we
simulate data drawn from our model light curves and compare various
scenarios.  One measure of the significance of the departures is to
simulate the data with all rotational effects and to compare with
model light curves ignoring either Kerr effects or time delay induced
perturbations.  For example, for our $P=3$ms fiducial light curves, we
find that pulse profiles depart significantly (95\% confidence K-S
statistic) from Doppler-alone models when the energy-summed total
light curve contains $3\times10^4$ counts. 
This does assume a simple model for the surface $T$ distribution
(although gravitational focusing renders results surprisingly
insensitive to the detailed cap shape).  To isolate rotational
perturbations in the presence of an arbitrary cap shape and $T$
profile would require significantly more data.  Table~\ref{tbl-1}
summarizes the counts required to see Doppler and
time-delay/frame-dragging induced distortions for various $P$. 

	Intra-band comparisons for a single pulsar are even more
robust, since cap shape and $T$ effects will be correlated at the
various energies. We have modeled the energy-dependent shape changes.
At $P=3$ms we find that a total of $6\times10^4$ counts are required
to measure the Doppler induced differences between narrow-band
lightcurves (10 bands assumed, logarithmically-spaced in energy).  For
a nearby millisecond pulsar with $T_{\rm cap, surf}=1.3 \times 10^6$K,
$N_H = 10^{19.5}$, we find the most statistically significant
differences are typically between the energies (at infinity) less than
0.14~keV and greater than 0.28~keV.  These estimates are conservative
in that we use only the pulse shape and phase differences.  The actual
count ratios between energy channels may be used to check the surface
$T$ dependence or to allow detection with even lower total counts. 

	We conclude that high precision models of pulsars such as
J0437-4715 must include at least Doppler boosting at the surface. For
example \citet{zav98} have published ROSAT PSPC light curves with good
statistics in energy bands extending to $1.1-2.0$~keV with $36\arcdeg$
binning.  Using model atmospheres and Schwarzschild photon
propagation, they have fit for $\alpha$ and $\zeta$, cap size and the
detailed $T$ distribution.  Our simulations (Fig.~\ref{all}) suggest
that the model light curves (and the parameter values fit) ignoring
rotational effects may be substantially in error. For example, fitted
$M/R_*$ ratios must be considered preliminary until the effects of
rotation are checked.  On the other hand the (formally independent)
$R_*$ and $M$ dependencies that can be traced to Doppler and
gravitational effects suggest that more precise tests of the EOS could
be effected with very high quality data and detailed models.  High
precision {\it XMM} and {\it AXAF} data, which are expected, should be
suitable for such fitting. 

	A second example is the newly-discovered 2.49ms
accretion-powered pulsar J1808-3658. This source displays strong
``soft lags'' \citep{wij98,cui98} in the light curves. \citet{for99}
has already suggested that this pulsar, as well as a number of QPO
sources, exhibits significant Doppler boosting which causes the high
energy pulse to lead the softer emission.  Alternative explanations,
such as Compton down-scatter in a soft corona have been suggested;
however, the energy-dependent pulse shape distortions \citep{cui98}
appear very similar to those of Fig.~\ref{energy}.  Our computations
thus provide significant support for the Doppler boosting
interpretation.  With more careful modeling it may possible to detect
the effect of other rotational distortions to the light curves of this
source. 

Doppler boosting effects are very significant for these short period
thermal cap emitters.  Time-delay and frame-dragging distortions,
although more subtle, have characteristic shapes that may allow
detection.  Sources with sharp pulse features, such as the radio and
magnetospheric X-ray millisecond pulsars, provide other interesting
targets. Resonant Compton scattering can also impart sharp light curve
features relatively high in the magnetosphere \citep{raj97}.  Doppler
effects will, of course, be larger at these altitudes, but significant
flux is ``lensed'' near the star surface in the light curves meaning
that the higher-order effects (time-delay and Kerr) may also be
significantly enhanced.  Modeling of such effects requires detailed
treatment of the curved space, swept-back dipole field structure. We
will describe such models in a future publication. 

\acknowledgments 

This work was supported in part by grants from NASA (NAG 5-3263) and
from the Research Corporation.  We thank the referee, George Pavlov,
for a very careful reading of the manuscript and a number of helpful
suggestions.

\end{document}